\documentclass[aps,prl,twocolumn,superscriptaddress]{revtex4-2}
\usepackage{amsmath,amssymb,graphicx,bm,xcolor}
\usepackage[colorlinks=true, linkcolor=blue, citecolor=blue, urlcolor=blue]{hyperref}

\begin{document}

\title{Curvature-Induced Magnon Frequency Combs}
\author{Hao Zhao}
\author{Qianjun Zheng}
\author{Peng Yan}
\email[Contact author: ]{yan@uestc.edu.cn}
\affiliation{School of Physics and State Key Laboratory of Electronic Thin Films and Integrated Devices, University of Electronic Science and Technology of China, Chengdu 610054, China}
%\affiliation{Institute of Fundamental and Frontier Sciences, Key Laboratory of Quantum Physics and Photonic Quantum Information of the Ministry of Education,University of Electronic Science and Technology of China, Chengdu 611731, China}

%\date{\today}

\begin{abstract}
Generating magnon frequency combs (MFCs) with tunable spacing via a single-frequency driving is crucial for practical applications but it typically relies on complex spin textures like skyrmions or vortices. Here, we theoretically and numerically demonstrate MFC generation in geometrically curved ferromagnetic thin films using single-frequency microwave excitation, without topological spin textures. We first show that the curvature transforms the planar ferromagnetic resonance into a localized, redshifted magnon bound state, which, under non-resonant driving, activates sequential three-magnon scattering processes assisted by the curvature-driven effective anisotropy and Dzyaloshinskii–Moriya interaction. It finally produces equally spaced, robust frequency combs with spacing exactly set by the bound mode frequency. Moreover, we find that the curvature gradient at the hybrid interface mimics an analog event horizon, with the bound state’s redshift resembling gravitational effects in black hole physics. Micromagnetic simulations confirm these curvature-driven nonlinear phenomenon, unveiling a novel geometric strategy for controlling magnon interactions and advancing compact magnonic devices.
\end{abstract}

\maketitle

\textit{Introduction}---Geometric curvature profoundly influences the physical properties of diverse systems, including optics \cite{roddier1988curvature,hocht1999seismics,schultheiss2010optics,cheneau2008geometric,
leonhardt2009transformation,kamien2009extrinsic,szameit2010geometric,wei2017negative,
salceda2015compact,spittel2015curvature}, electronics \cite{ferrari2008schrodinger,ortix2010effect,atanasov2010tuning,onoe2012observation,wang2014pauli,gomes2020electronic,zhao2021influence,gentile2022electronic}, thermodynamics \cite{ruppeiner1995riemannian,ruppeiner2008thermodynamic,konig2004morphological,williams1991thermodynamics,an2025curvature}, and liquid crystals \cite{frank1958liquid,ondris1993curvature,shri1996curvature,kunieda1998effect,santangelo2005curvature,serra2016curvature,mostajeran2016encoding,kowalski2018curvature}. In magnetic systems, curvature effects have been theoretically analyzed \cite{Gaididei2014,Streubel2016,sheka2015curvature,Yershov2023Control} and experimentally validated \cite{streubel2012magnetically,volkov2019experimental,turvcan2021spin,korber2021symmetry}, fueling extensive research \cite{pylypovskyi2020curvilinear,sheka2021perspective,bittencourt2021curvature,edstrom2022curved,zhao2023long,Salamone2024Interface,Yershov2025Curvature}. The interplay of geometry and magnetism induces effective Dzyaloshinskii-Moriya interactions (DMI) and curvature-driven anisotropies, stabilizing exotic magnetic textures such as vortices and skyrmions \cite{Vojkovic2017VortexAntivortex,Wyss2017ImagingMagneticVortex,Kravchuk2016,Kravchuk2018MultipletSkyrmion}. Beyond static textures, curvature modifies spin-wave propagation, enabling non-reciprocal dispersion and tunable magnon (the quantum of spin wave) modes \cite{gaididei2018localization,Otalora2016,otalora2017asymmetric,wu2022curvilinear}. Notably, curvilinear magnonics also offers a platform for analog gravity, simulating black-hole-like phenomena such as event horizons in magnon dynamics \cite{Skyba2019,Errani2025}.

While linear and static curvature effects are well-explored, the impact of curvature on nonlinear magnon dynamics remains largely uncharted. In planar magnets, single-mode driven magnon frequency combs (MFCs), equidistant coherent spectral peaks crucial for applications, require interactions with topological textures like skyrmions or vortices \cite{Wang2021Magnonic,wang2022twisted,liang2024asymmetric,Hens2025}. A critical question is whether MFCs can emerge purely from geometric curvature, without such textures or material patterning. This inquiry is amplified by analog gravity models, where curvature gradients mimic event horizons, potentially enhancing nonlinear wave interactions via mechanisms reminiscent of Hawking radiation or superradiance \cite{Skyba2019,Errani2025}.

In this Letter, we demonstrate that geometric curvature alone can induce magnon bound state and trigger three-magnon interactions between localized and propagating modes, generating robust MFCs under \emph{single-frequency} microwave driving, without topological spin textures. We analyze a rotationally symmetric curved surface with tunable curvature, which reshapes the ferromagnetic ground state, transforming the ferromagnetic resonance (FMR) into a bound state with a redshifted frequency. Assited by the curvature-induced effective anisotropy and DMI, this mode drives efficient three-magnon scattering, enabling the formation of MFC. Micromagnetic simulations validate these predictions, showing robust MFCs emerging without solitons or engineered defects. Finally, we point out that the curvature gradient at the hybrid junction mimics an analog event horizon, where redshifted modes and Hawking-like fluctuations amplify comb formation, forging compelling parallels to black hole physics.

\begin{figure}[t]
        \centering
        \includegraphics[width=0.49\textwidth]{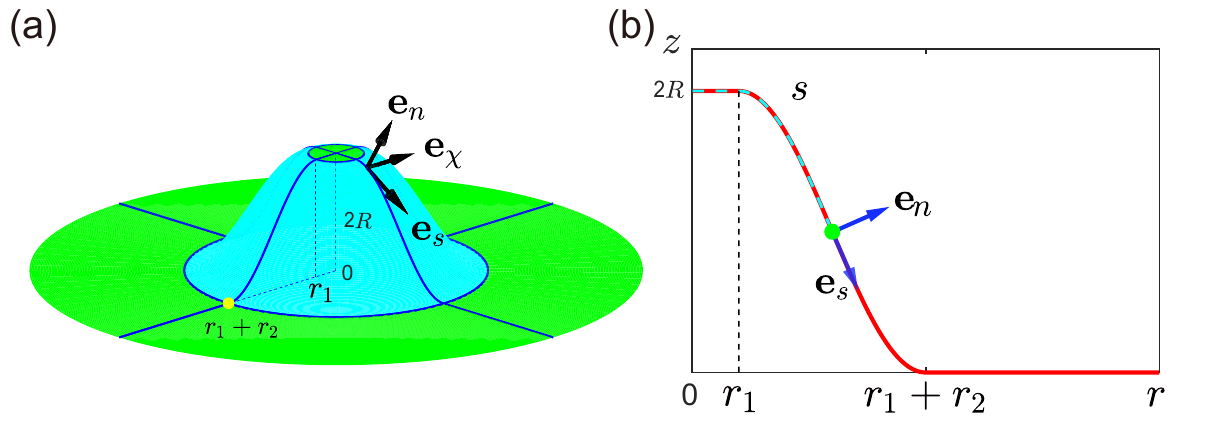}
        \caption{(a) Rotationally symmetric surface generated by revolving the curve in (b) about the $z$-axis. Black arrows depict the local orthonormal basis $(\mathbf{e}_s, \mathbf{e}_\chi, \mathbf{e}_n)$ of the curvilinear coordinate system. (b) Profile of the smooth curve $z(r)$ in the $r$-$z$ plane, with the central section ($r_1\leqslant r \leqslant r_1+r_2$) defined by a half-period cosine function of amplitude $R$. Arrows indicate the tangential and normal unit vectors at a point on the curve. The dashed cyan line represents the arc length $s$ measured from $r=0$.}
        \label{fig1}
\end{figure}
\textit{Model}---Let's start with a general curved thin ferromagnetic film, with thickness $L$ being sufficiently small to ensure uniform magnetization across the thickness. To describe magnetic moments on this surface, we use a curvilinear orthonormal basis $(\mathbf{e}_s, \mathbf{e}_\chi, \mathbf{e}_n)$, representing meridional, azimuthal, and normal directions, respectively, as shown in Fig. \ref{fig1}. The total energy includes exchange and perpendicular anisotropy contributions
\begin{equation}
\mathcal{E} =
\mu_0 M_s L \int \left( A \mathcal{E}_{\text{ex}} - K m_n^2 \right) d\mathbf{r},
\end{equation}
where $\mu_0$ is the vacuum permeability, $M_s$ is the saturation magnetization, $A$ is the exchange stiffness, $K > 0$ is the easy-normal anisotropy constant, and $\mathbf{m} = (m_s, m_{\chi}, m_n)$ is the unit magnetization vector, expressed as $ m_s = \sin\theta \cos\psi$, $m_{\chi} = \sin\theta \sin\psi$,  $m_n = \cos\theta$, with $\theta$ and $\psi$ denoting the polar and azimuthal angles in the local frame, respectively.
The exchange energy density is \cite{Gaididei2014,sheka2015curvature}
\begin{equation}
\mathcal{E}_{\text{ex}} = (\bm{\nabla}\theta - \boldsymbol{\Gamma})^2 + [\sin\theta(\bm{\nabla}\psi - \boldsymbol{\Omega}) - \cos\theta\, \partial_\psi \boldsymbol{\Gamma} ]^2,
\label{eq:exchange}
\end{equation}
where $\bm{\Omega}$ is the modified spin connection and $\bm{\Gamma}$ is the geometric vector potential (see Supplemental Material Sec. I \cite{supplemental}). From Eq.~(\ref{eq:exchange}), one can observe that $(\bm{\nabla}\theta)^2+\sin^2\theta(\bm{\nabla}\psi)^2$ takes the same form as the exchange energy in a planar geometry. The couplings of $\bm{\nabla}\theta$ and $\bm{\nabla}\psi$ with $\bm{\Gamma}$ and $\bm{\Omega}$, e.g., $\bm{\Gamma}\cdot\bm{\nabla}\theta$, $\sin^2\theta\,\bm{\Omega}\cdot\bm{\nabla}\psi$, and $\sin\theta\cos\theta\,\bm{\Omega}\cdot \partial_\psi \boldsymbol{\Gamma} $, generate the effective DMI, while the rest terms constitute the effective anisotropy. 

For the sake of analytical derivation, we consider a surface generated by revolving a smooth curve $z(r)$ about the $z$-axis, defined as
\begin{equation}
    z(r)=
    \begin{cases}
     2R, &r \leqslant r_1,\\
     R\big[1+\cos\frac{\pi(r-r_1)}{r_2}\big], & r_1<r \leqslant r_1+r_2,\\
     0,& r>r_1+r_2,
     \end{cases}
\end{equation}
where $2R$ is the plateau height, $r$ is the radial coordinate, and the cosine profile ensures a smooth transition between the curved region ($r_1<r \leqslant r_1+r_2$), plateau ($r \leqslant r_1$), and flat exterior ($r>r_1+r_2$). Then we obtain $ \boldsymbol{\Gamma} = \varkappa_1 \cos\psi\, \mathbf{e}_s + \varkappa_2 \sin\psi\, \mathbf{e}_\chi,$ and $\quad \boldsymbol{\Omega} = -\mathbf{e}_\chi\, \frac{r'}{r}$, where $ \varkappa_1=\frac{z''}{r'}$, $\varkappa_2=\frac{z'}{r}$, and primes denote derivatives with respect to the arc-length $s$, e.g., $r'=\partial_{s}r$. In numerical calculations, we adopt material parameters of yttrium iron garnet (YIG), and choose $r_1=10\ \text{nm}$ and $r_2=50\ \text{nm}$ (see below).

Due to the rotational symmetry ($\psi=0$), the static polar angle $\theta=\Theta$ satisfies
\begin{equation}
 \frac{\left( r \Theta' \right)'}{r}-\sin\Theta\cos\Theta \Xi-2\frac{r'}{r}\varkappa_2\sin^2\Theta=\varkappa_1'+\varkappa_2',
\end{equation}
where $ \Xi = \frac{K}{A} + \left( \frac{r'}{r} \right)^2 - \varkappa_2^2$. 
\begin{figure}[t]
        \includegraphics[width=0.49\textwidth]{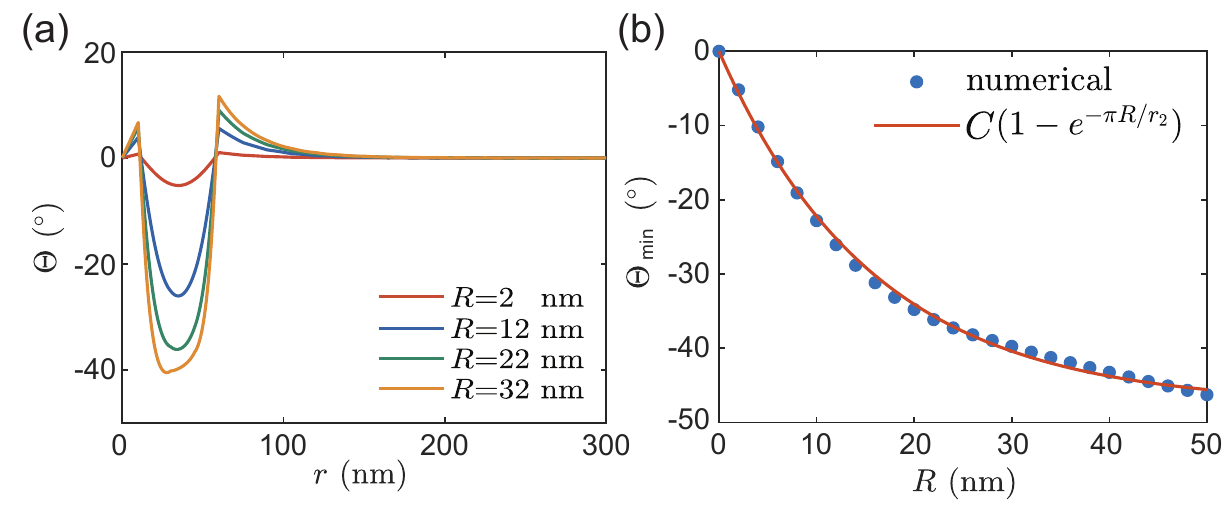}
        \caption{(a) Radial distribution of the ground-state polar angle $\Theta$ for various surface heights $R$. (b) Minimum polar angle $\Theta_{\min}$ versus $R$, with symbols from numerical calculations and red curve from analytical formula.}
        \label{fig2}
\end{figure}
To ensure regularity at $r = 0$, we expand $\Theta(r) \approx \lambda_1 r + \lambda_2 r^2 + \cdots$, and impose $\Theta(r \to \infty)=0$ for asymptotic flatness. This yields a ground-state profile $\Theta(r)$ driven purely by curvature.

Figure \ref{fig2}(a) shows that the curvature-induced deformation intensifies with increasing surface height $R$. The minimum polar angle $\Theta_{\min}$ follows $\Theta_{\min} \approx C (1 - e^{-\pi R / r_2})$, with a fitting parameter $C = -0.8314$, see solid red curve in Fig. \ref{fig2}(b).

\textit{Curvature-induced magnon bound state}---To investigate magnon excitations around the curvature-induced ground state $\Theta(r)$, we introduce $\theta = \Theta + \vartheta$ and $\psi=\Phi+\phi/\sin\Theta$, with $\Phi=0$ the static solution of $\psi$. Substituting them into the damping-free Landau-Lifshitz-Gilbert (LLG) equation, the linearization yields coupled wave equations \cite{Kravchuk2018MultipletSkyrmion}
\begin{equation}
\begin{aligned}
-\frac{\partial_t \vartheta}{2\gamma A} &= -\nabla^2 \phi - U_1(r)  \phi - W(r)\, \partial_\chi \vartheta,\\
\frac{\partial_t \phi}{2\gamma A} &= -\nabla^2 \vartheta + U_2(r)  \vartheta + W(r) \, \partial_\chi \phi, 
\end{aligned}
\label{eq:LLGlinear}
\end{equation}
where the potentials are
\begin{equation}
\begin{aligned}
U_1(r)  &= \cos^2\Theta\, \Xi - \Theta'^2 - \varkappa_1^2 + \varkappa_2^2 + 2\varkappa_1 \Theta' + \frac{\varkappa_2 r'}{r} \sin(2\Theta), \\
U_2(r)  &= \cos(2\Theta)\, \Xi + \frac{2\varkappa_2 r'}{r} \sin(2\Theta), \\
W(r)  &= \frac{2r'}{r^2} \cos\Theta + \frac{2\varkappa_2}{r} \sin\Theta.
\end{aligned}
\label{eq:UVW}
\end{equation}

We seek solutions of the form $\vartheta(r,\chi,t) = {\vartheta}(r)\cos(\omega t + m\chi) $ and $ \phi(r,\chi,t) = {\phi}(r)\sin(\omega t + m\chi ) $, where $\omega$ is the magnon frequency, $ m \in \mathbb{Z} $ is the azimuthal quantum number, and $\chi$ is the azimuthal angle. The system reformulates as a Schrödinger-like eigenvalue problem
\begin{equation}
\mathcal{H}\, \Phi = \frac{\omega}{\gamma A}\, \sigma_x\, \Phi, \quad \Phi = 
\begin{pmatrix}
{\vartheta} \\
{\phi}
\end{pmatrix},
\label{eq:schrodinger}
\end{equation}
with
\[
\mathcal{H} = 
\begin{pmatrix}
-\nabla_r^2 + \frac{m^2}{r^2} + U_2(r) & m W(r)  \\
m W(r)  & -\nabla_r^2 + \frac{m^2}{r^2} + U_1(r) 
\end{pmatrix},
\]
where $ \nabla_r^2 = \frac{r'}{r} \partial_r (rr' \partial_r) $ is the radial Laplacian and $\sigma_x$ is the Pauli matrix.

For small surface height $R$, the differences between $U_1(r)$ and $U_2(r)$, and between $\vartheta$ and $\phi$, are negligible. Equation~(\ref{eq:schrodinger}) then can be simplified to
\begin{equation}
    \nabla_r^2 \xi+  U_{\text{eff}}(r) \xi=\frac{\omega}{2\gamma A} \xi,
\end{equation}
where
\begin{equation}
U_{\text{eff}}(r) = m W + \frac{m^2}{r^2} + \frac{1}{2}(U_1 + U_2),
\label{eq:Ueff}
\end{equation}
and $\xi=(\vartheta+\phi)/2$.

\begin{figure}[t]
        \includegraphics[width=0.49\textwidth]{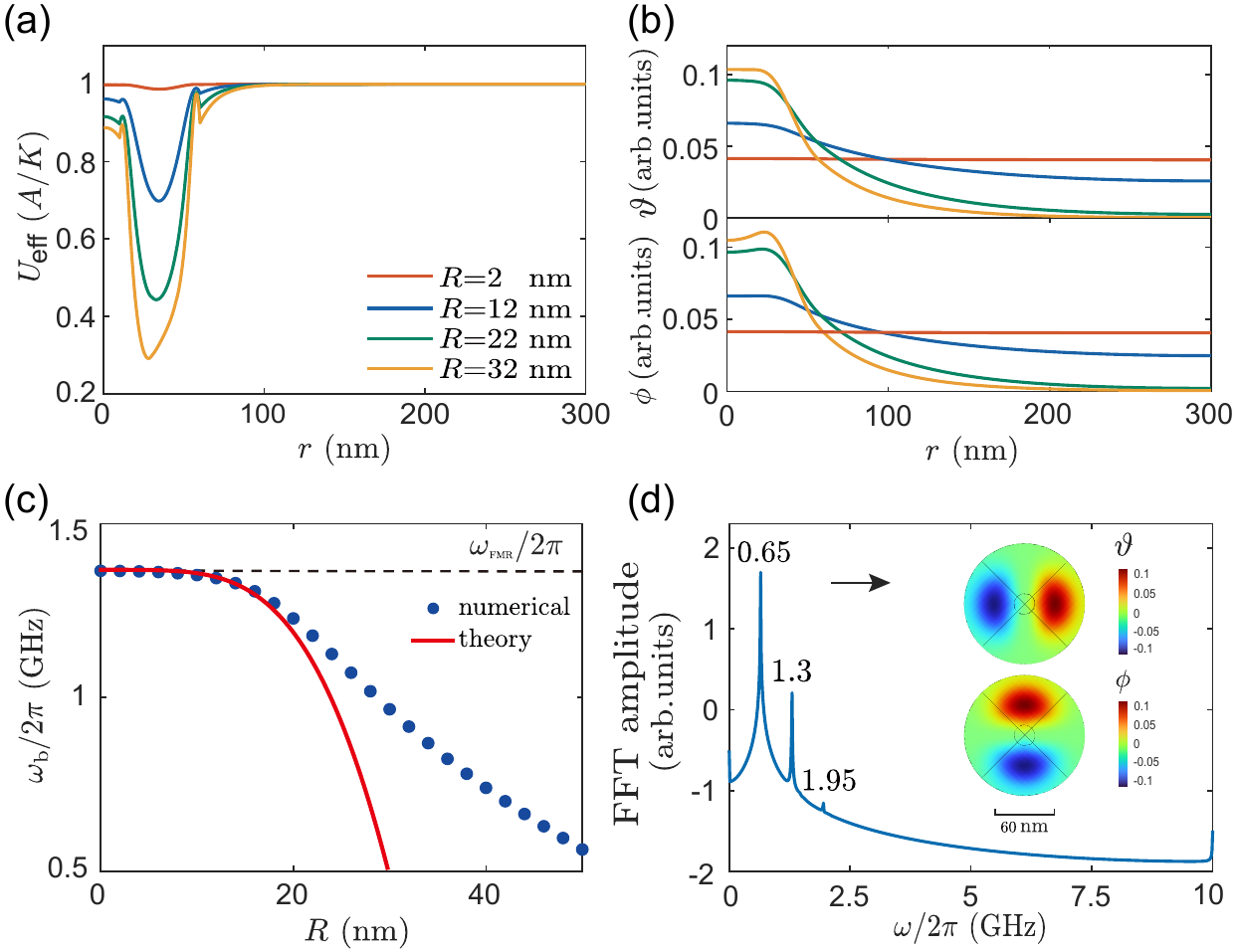}
        \caption{(a) Spatial distribution of the effective potential $U_{\text{eff}}$ for various surface heights. (b) Profile of bound magnon states for different $R$. (c) Frequency of the curvature-induced magnon bound state versus the surface height, with symbols from numerical simulation and curve from analytical formula (\ref{eq:Bound}). (d) Magnon spectrum for $R = 45$ nm under a sinc-function field $h_0 \text{sinc}(\omega_c t)\hat{x}$ over the curved region ($r\leqslant r_1+r_2$) with cutoff frequency $\omega_c / 2\pi = 10$ GHz and amplitude $\mu_0 h_0 = 10$ mT. Inset: distribution of $\vartheta$ and $\phi$ in the $r$-$\chi$ plane for the bounded magnon mode at $0.65\ \text{GHz}$.}
        \label{fig3}
\end{figure}
For the minimal model with $m = -1$, the effective potential satisfies $U_{\text{eff}}(0)=\frac{K}{A}-\frac{\Theta'^2}{2}$ at the surface center and $U_{\text{eff}}(\infty)=\frac{K}{A}$ at infinity. It is noted that $U_{\text{eff}}(0) < U_{\text{eff}}(\infty)$, which thus forms a potential well [Fig. \ref{fig3}(a)]. Additionally, we observe that the minima of $U_{\text{eff}}(r)$ and $\Theta(r)$ coincide. This yields $U_{\text{eff},\min}\approx\frac{1}{2}(\cos^2\Theta_{\min}+\cos2\Theta_{\min})(\frac{K}{A}-\varkappa_{2,c}^2)+\varkappa_{2,c}^2+3\frac{\varkappa_{2,c} r'_c}{r_c} \sin2\Theta_{\min}-4\frac{\varkappa_2}{r_c} \sin\Theta_{\min}$, where $r_c=r_1+\frac{r_2}{2}$, and $\varkappa_{2,c}=\varkappa_{2}(r_c)$. Increasing $R$ deepens the well, and thus enhances the mode localization, as shown Fig. \ref{fig3}(b). 

Next, we treat $U_{\text{eff}}(r)$ as a cosine potential in the small height limit $\pi R\ll r_2 $. We then derive the bound state frequency $\omega_b \approx \omega_{\text{FMR}}-\delta\omega$, where $\omega_{\text{FMR}}=2\gamma K$ is the FMR angular frequency and $\delta\omega=\frac{\gamma AU_d^2 w^2}{2\pi^2}$, with $U_d=\frac{K}{A}-U_{\text{eff},\min}$ the well depth and $w=\frac{2r_2}{\pi}E(\frac{q-1}{q})\sqrt{q}$ being the curved region's arc length. Here, $q=1+(\frac{\pi R}{r_2})^2$ and $E(y)$ is the complete elliptic integral of the second kind. Up to fourth-order, we obtain (Sec. II \cite{supplemental})
\begin{equation}
\omega_b=2\gamma K-\frac{\gamma \pi^2(4C A+3C^2 r_c^2 K)^2}{8 A r_2^2 r_c^4}  R^4.
\label{eq:Bound}
\end{equation} 

Figure \ref{fig3}(c) plots the bound state frequency $\omega_b$ as a function of the surface height $R$, showing that a deeper well allows a smaller frequency. Our theoretical formula (\ref{eq:Bound}) agrees well with numerical results in the small $R$ region. For $m \neq -1$, the centrifugal barrier prevents localization, indicating that only the $m = -1$ mode is geometrically bounded. Numerical simulations validate these results, showing a low-frequency bound state with a profile matching theory [see Fig. \ref{fig3}(d)]. Meanwhile, we have observed two higher-frequency modes at 1.3 and 1.95 GHz, corresponding to the second and third harmonics of the fundamental magnon bound state, respectively.

\textit{MFCs}---To analytically study nonlinear magnon-magnon interactions, we adopt the vectorial Hamiltonian formalism \cite{tyberkevych2020vector,verba2021}, expressing the spatiotemporal magnetization vector $\mathbf{m}(\mathbf{r},t)$ as
\begin{equation}\label{eq-1-2}
\begin{aligned}
    \mathbf{m}(\mathbf{r},t)= \big[1-\frac{\mathbf{s}^2(\mathbf{r},t)}{2}\big]\bm{\mu}(\mathbf{r})+\sqrt{1-\frac{\mathbf{s}^2(\mathbf{r},t)}{4}}\mathbf{s}(\mathbf{r},t),
\end{aligned}    
\end{equation}
where $\bm{\mu}(\mathbf{r}) = \mathbf{m}_0(\mathbf{r})$ is the normalized static magnetization, and $\mathbf{s}(\mathbf{r},t)$ is the dimensionless dynamic magnetization, perpendicular to $\bm{\mu}$ at each point (i.e., $\mathbf{s} \perp \bm{\mu}$). We then expand the dynamic magnetization in magnon eigenmodes $\mathbf{s}_\nu(\mathbf{r})$
\begin{equation}\label{eq-1-4}
    \mathbf{s}(\mathbf{r},t)=\sum_\nu \big[c_\nu(t) \mathbf{s}_\nu(\mathbf{r})+c.c\big],
\end{equation}
with $c_\nu(t)$ as the time-dependent complex amplitudes. The Hamiltonian for three-magnon processes, describing the confluence of magnons 1 and 2 into magnon 3 (and the reverse splitting), is
\begin{equation}
    \mathcal{H}_{3}=\sum_{123} \bar{\mathcal{V}}_{12,3}c_1 c_2 c_3^*+c.c,
\end{equation}
where $\bar{\mathcal{V}}_{12,3}$ is the interaction vertex (detailed in Sec. III \cite{supplemental}).

It is noted that the bound magnon mode $c_r$ serves as a pivotal mediator for nonlinear magnon interactions. Coupled with incident magnons $c_k$ excited by external microwaves, it enables sum- and difference-frequency generations (i.e., $c_p$ and $c_q$ modes), ultimately producing an MFC. The Heisenberg equations for these operators are
\begin{equation}
\begin{aligned}
&i\frac{dc_k}{dt}=(\omega_k-i\alpha \omega_k)c_k+\bar{\mathcal{V}}_{rq,k}^* c_r c_q+\bar{\mathcal{V}}_{rk,p}c^*_rc_p+h e^{i \omega_d t},\\
&i\frac{dc_r}{dt}=(\omega_r-i\alpha \omega_r)c_r+\bar{\mathcal{V}}_{rq,k}^* c_kc_q^*+\bar{\mathcal{V}}_{rk,p}c^*_kc_p,\\
&i\frac{dc_p}{dt}=(\omega_p-i\alpha \omega_p)c_p+\bar{\mathcal{V}}_{rk,p}c_kc_r,\\
&i\frac{dc_q}{dt}=(\omega_q-i\alpha \omega_q)c_q+\bar{\mathcal{V}}_{rq,k}^* c_kc_r^*,
\end{aligned}
\label{eq:Numerical}
\end{equation}
where $\omega_\nu\ (\nu = k, r, p, q)$ is the mode frequency, $\omega_d$ is the driving frequency (matching $\omega_k$), $\alpha$ is the Gilbert damping constant, and $h$ is the driving filed amplitude.
\begin{figure}[t]
        \includegraphics[width=0.49\textwidth]{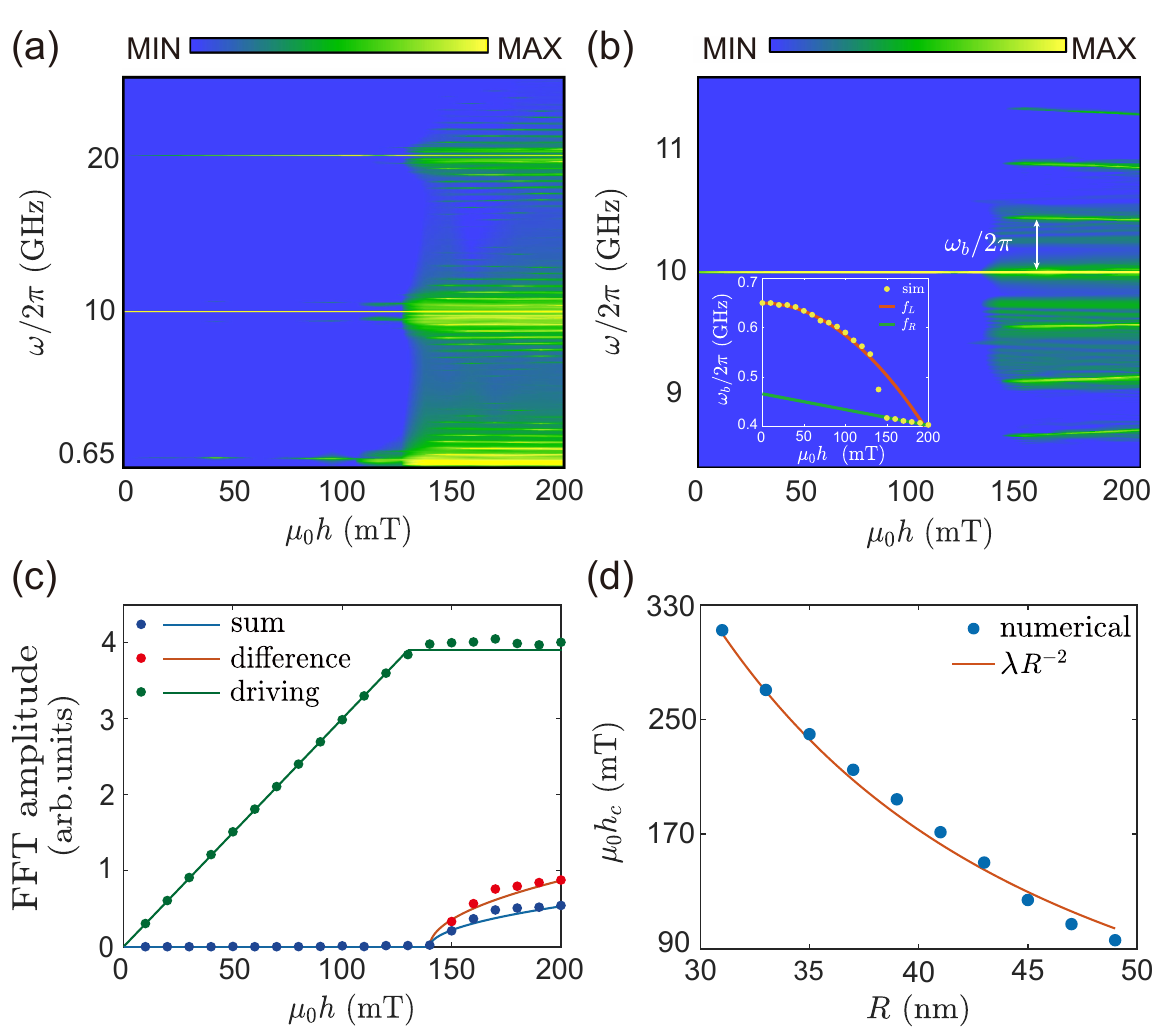}
        \caption{(a) System response versus driving amplitude $\mu_0 h$ for fixed $R = 45$ nm; color indicates excitation amplitude. (b) Enlarged view near 10 GHz from (a). Inset: frequency of the magnon bound state as a function of the drive-field amplitude. (c) Driving-, sum- and difference-frequency peaks amplitudes versus driving field strength; inflection points indicate a 140 mT threshold, with dots from micromagnetic simulations, and curves from solutions of Eq. \eqref{eq:Numerical}. (d) Threshold field versus surface height $R$: Blue points from simulations, red line from analytical fitting. }
        \label{fig4}
\end{figure}

To verify the emergence of MFC, we conduct full micromagnetic simulations using $\texttt{COMSOL}$ micromagnetic simulation module \cite{Weichao2023}. A sinusoidal magnetic field $\mathbf{h}=h\sin(\omega_d t)\hat{x}$ with $\omega_d/2\pi=10$ GHz is applied over the region $r\leqslant r_1+r_2$ to excite magnons. YIG parameters are used: saturation magnetization $M_s=1.94\times 10^{5}\ \text{A}/\text{m}$, exchange stiffness $A = 1.64 \times 10^{-11}\ \text{Am}$, gyromagnetic ratio $\gamma = 2.21\times 10^5$ (rad$/$s)$/(\text{A}/\text{m})$, Gilbert damping $\alpha = 1 \times 10^{-4}$, and perpendicular anisotropy $K= 1.94 \times 10^4\ \text{A}/\text{m}$. We analyze the spertum by performing fast Fourier transformation (FFT) of $m_\chi$. As shown in Fig. \ref{fig4}(a), the spectrum initially displays a single peak at the driving frequency and its multiplications in the low-field region ($\mu_0h<100\ \text{mT}$). Increasing the microwave amplitude generates multiple sidebands from three-magnon processes, forming a regular comb ($\mu_0h>150\ \text{mT}$). A secondary comb appears around the second harmonic (20 GHz). Figure \ref{fig4}(b) details the MFC near 10 GHz, spaced by $\omega_b$, showing that curvature-induced bound modes enhance the nonlinear magnon-magnon scattering, enabling robust MFCs without textures or resonators. Slanted lines indicate the microwave detuning of the bound magnon frequency and thus the comb spacing. The inset in Fig. \ref{fig4}(b) plots $\omega_b$ as a function of the driving-field amplitude $h$. It shows a quadratic (linear) dependence on $h$ below (above) the threshold field, where $f_L=\frac{\omega_b}{2\pi}-\beta\mu_0^2h^2$ (red curve) and $f_R=\frac{\omega_b}{2\pi}-\epsilon \mu_0^2 h_c^2-\rho\mu_0 h$ (green line) with fitting coefficients $\beta=6.4\ \text{GHz}/\text{T}^2$, $\epsilon=9.4\ \text{GHz}/\text{T}^2$, and $\rho=0.325\ \text{GHz}/\text{T}$. Here, $h_c$ is the threshold microwave field, as analyzed below. We attribute the exotic frequency shift to the four-magnon interaction, since the shift value scales linearly with the magnon occupation number (see Sec. VI \cite{supplemental}). Figure \ref{fig4}(c) shows a good agreement between our theoretical model (curves) and micromagnetic simulations (symbols).

The onset of the MFC is marked by a threshold field $h_c$, above which nonlinear interactions dominate.  When $\omega_r \ll \omega_k$ and $\bar{\mathcal{V}}_{rq,k}^* \approx \bar{\mathcal{V}}_{rk,p} = \bar{\mathcal{V}}_3$, the threshold is approximately
\begin{equation}
h_c \approx \frac{\alpha^2 \omega_k}{\sqrt{2} \mu_0\bar{\mathcal{V}}_3},
\end{equation} indicating that $h_c$ is inversely proportional to $\bar{\mathcal{V}}_3$. For small curvature, the three-magnon vertex $\bar{\mathcal{V}}_3$ scales linearly with $R$, arising from the curvature-driven effective DMI \cite{supplemental}. At large curvature, the dominant contribution originates from the quadratic terms in the self-interaction tensor $\hat{\bm{N}}_a$ \cite{supplemental}, due to the anisotropy---including both the intrinsic easy-normal anisotropy and the curvature-driven effective anisotropy, yielding $\bar{\mathcal{V}}_3 \propto R^2$, and thus $h_c \propto R^{-2}$. Simulation results for $h_c$, shown in Fig. \ref{fig4}(d), closely align with this prediction with a fitting parameter $\lambda=2.936\times10^{-16}\ \text{m}^2\text{T}$.

\textit{Discussion}---While our analysis employs a cosine-shaped film, the results are applicable to generic geometries like torus and sphere, provided a sufficient curvature gradient exists to localize magnons. In addition, the curvatures in our hybrid structure establish compelling parallels with black hole physics in general relativity (GR), particularly through analog gravity effects in curvilinear magnonics \cite{Skyba2019,Errani2025}. The junction between the curved magnet and flat film functions as an analog event horizon, where the abrupt curvature gradient $\varkappa = \pi^{2}R/r^{2}_2$ creates a potential barrier for magnons, resulting in mode redshift and amplification reminiscent of gravitational effects near black holes. The magnon frequency redshift $\zeta = \omega_b / \omega_{\text{FMR}} \approx \sqrt{1 - (\varkappa \ell)^4/16}$, with magnetic length $\ell =\sqrt{\frac{-\sqrt{2} \pi(4C A+3C^2 r_c^2 K)}{ \sqrt{A K} r_2}}\frac{r_2^2}{\pi^2 r_c}$ determined by Eq. \eqref{eq:Bound}, quantifies the frequency reduction of bound modes with increasing curvature, approaching zero at high values. This quartic scaling suggests a strong-field regime, differing from the weak-field GR redshift but aligning with higher-order post-Newtonian corrections \cite{Gravitation}. Finally, we note that the junction enables analog Hawking-like fluctuations with effective temperature $T_H = \hbar v_m / (8\pi k_B\ell)$, where $v_m$ is the magnon velocity. For $R = 45\ \text{nm}$, we have $\ell=6.966\ \text{nm}$. Assuming $v_{m}\approx10^{3}\ \text{m}/\text{s}$, one can estimate $T_H \approx 43.6\ \text{mK}$, providing minimal but conceptually significant noise that seeds MFC cascades, lowering thresholds in quantum-limited regimes, akin to stimulated Hawking radiation enhancing black hole spectra \cite{Hawking1974}.

\textit{Conclusion}---To summarize, we have demonstrated that the geometric curvature in ferromagnetic thin films induces a transition from a uniform ground state to a curvature-modulated magnetization profile, yielding a spatially localized magnon bound state distinct from planar FMR. This mode enables robust MFCs through enhanced nonlinear interactions under a single-frequency microwave driving, eliminating the need for solitonic textures or engineered resonators. The profound insight of our findings lies in harnessing curvature alone to achieve tunable, texture-free MFCs, with striking analogies to black hole physics through redshift and Hawking-like fluctuations. These results not only redefine magnon control strategies but also position curvature as a powerful tool for next-generation magnonic devices.
\begin{acknowledgments}
This work was funded by the National Key R$\&$D Program of China (No. 2022YFA1402802 and No. 2025YFA1411302), the National Natural
Science Foundation of China (NSFC) (No. 12374103
and No. 12434003), and Sichuan Science and Technology program (No. 2025NSFJQ0045).
\end{acknowledgments}
\nocite{*}

%\bibliography{ref.bib}

\end{document}